\documentclass[runningheads]{svmult}

\usepackage{makeidx}   % allows index generation
\usepackage{graphicx}  % standard LaTeX graphics tool
\usepackage{subeqnar}  % subnumbers individual equations
\usepackage{multicol}  % used for the two-column index
\usepackage{physprbb}  % modified textarea for proceedings,
\makeindex             % used for the subject index
\newcommand{\arcsecsgreve}{\mbox{$^{\prime\prime}$}}
\newcommand{\Lsolargreve}{\mbox{$L_{\odot}\,$}}
\newcommand{\lsgreve}{\mathrel{\raise0.35ex\hbox{$\scriptstyle <$}\kern-0.6em\lower0.40ex\hbox{{$\scriptstyle \sim$}}}}
\begin{document}
\title*{A 1200\,$\mu$m MAMBO Survey of the ELAIS\,N2 and Lockman Hole East Fields}
\toctitle{A 1200\,$\mu$m MAMBO Survey of the ELAIS\,N2 and Lockman Hole East Fields}
%\protect\newline in the Particle Deflection Plane}
% allows explicit linebreak for the table of content
%
%
\titlerunning{A 1200\,$\mu$m MAMBO Survey of the ELAIS\,N2 and Lockman Hole East Fields}
% allows abbreviation of title, if the full title is too long
% to fit in the running head
%
\author{T.~R.~Greve\inst{1}
\and R.~J.~Ivison\inst{1}
\and F.~Bertoldi\inst{2}
\and J.~A.~Stevens\inst{1}
\and S.~C.~Chapman\inst{3}
\and I.~Smail\inst{4}
\and A.~W.~Blain\inst{3}}
\authorrunning{Greve et al.}
% if there are more than two authors,
% please abbreviate author list for running head
%
%
\institute{Royal Observatory Edinburgh, Blackford Hill, Edinburgh EH9 3HJ, UK
%\and Astronomy Technology Centre, Royal Observatory, \\
%     Blackford Hill, Edinburgh EH9 3HJ, UK\\
\and Max-Planck Institut f\"{u}r Radioastronomie, Bonn, Germany\\
\and Astronomy Department, Caltech, Pasadena, CA 91125, USA\\
\and Department of Physics, University of Durham, South Road, Durham DH1 3LE, UK}

\maketitle              % typesets the title of the contribution

\begin{abstract}
Using the MPIfR Max Planck Millimeter Bolometer array (MAMBO) on the IRAM 30m Telescope
we have mapped the ELAIS\,N2 and Lockman Hole East Fields at 1200\,$\mu$m to a rms
noise level of $0.8-1.0$\,mJy per $11\arcsecsgreve$ beam. The areas surveyed
are 326 arcmin$^2$ in the ELAIS\,N2 field and 212 arcmin$^2$ in the Lockman Hole\footnote{The
Lockman data are part of the MAMBO 1sq.~deg.~survey (Bertoldi et al.~in prep.)}, and
cover the 260 arcmin$^2$ previously observed by SCUBA \cite{Scott-et-al-2002}.
\end{abstract}

\begin{figure}[h]
\begin{center}
\includegraphics[width=0.663\textwidth]{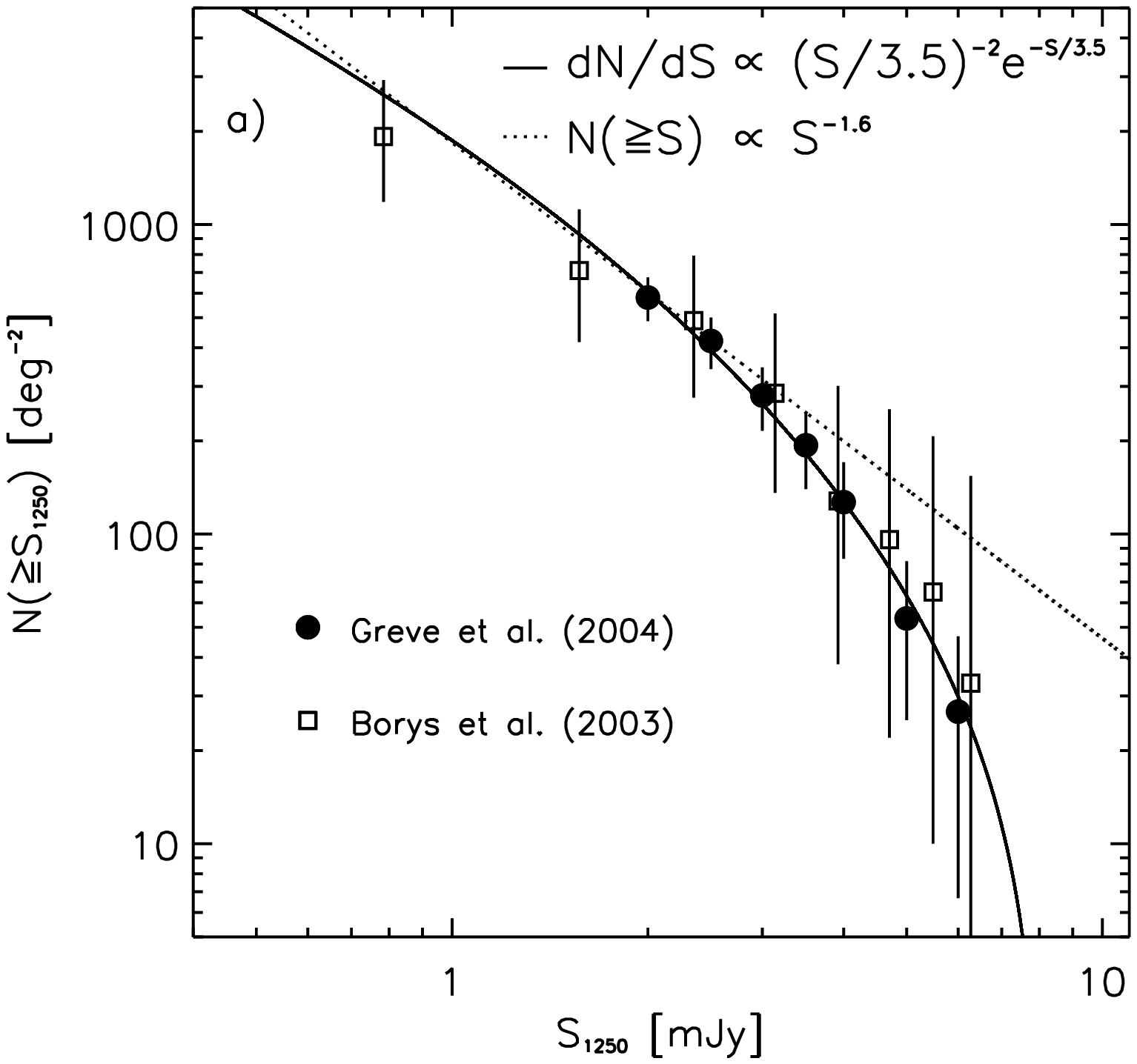}~~~~\includegraphics[width=0.3072\hsize, angle=0]{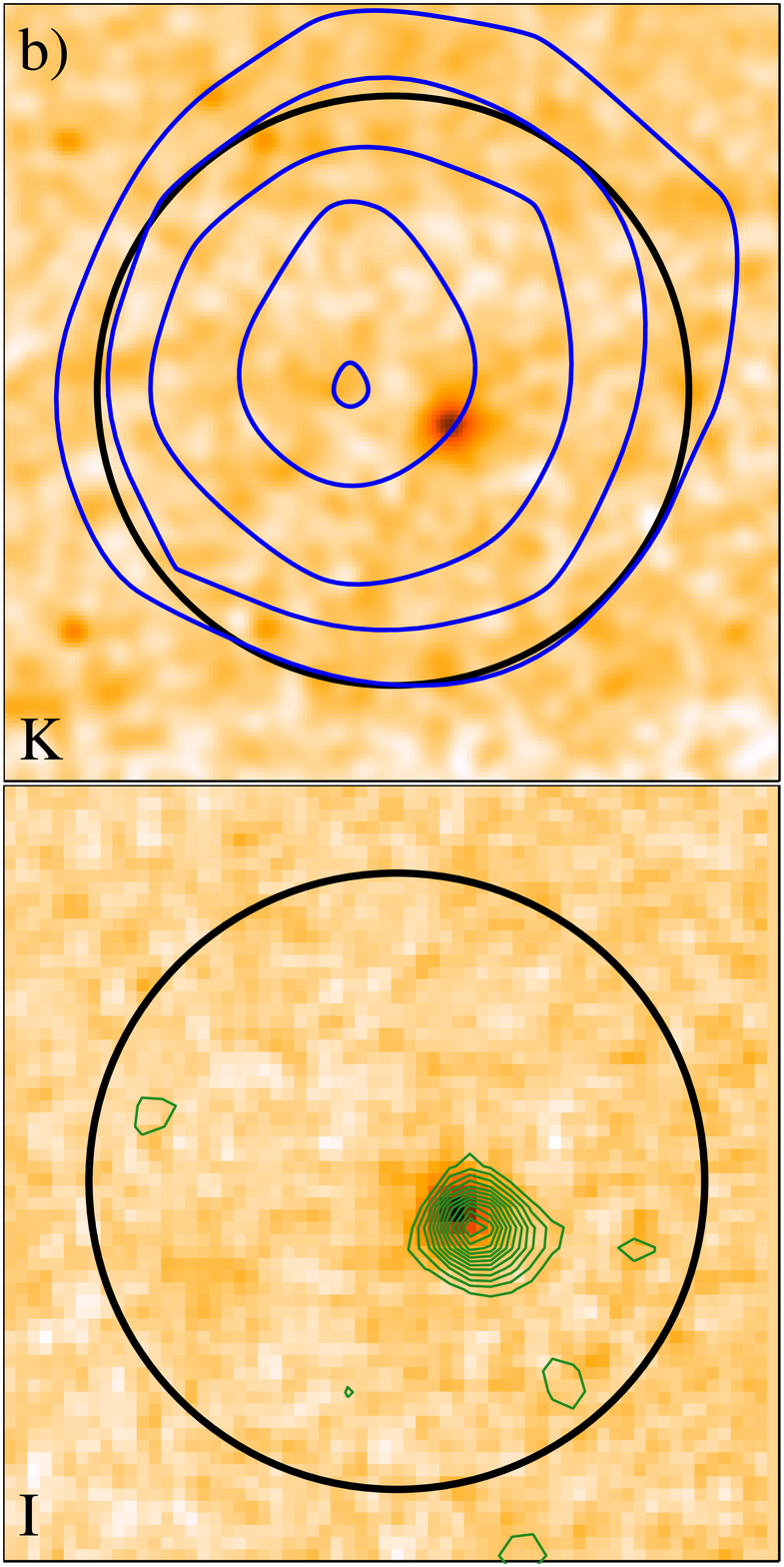}
\end{center}
\caption[Cumulative number counts at 1200\,$\mu$m.]{{\textbf a)} Preliminary cumulative number counts at 
1200\,$\mu$m (filled circles) based on $\geq 3.5\sigma$ sources extracted from our 
MAMBO map of the ELAIS\,N2 and Lockman Hole East fields. 
850\,$\mu$m counts based on the HDF-North SCUBA Super-map are shown as squares
(see Borys et al.~(2003) for details).
Note the 850\,$\mu$m fluxes have been scaled by a factor of $1/2.5 = 0.4$.
{\textbf b)} An example of a MAMBO source with a strong radio counterpart. {\it Top:} The 1200\,$\mu$m-emission shown as
blue contours: 3.5,4.0,4.5,5.0, 5.5$\times \sigma$ with $\sigma = 0.8$\,mJy;
{\it bottom:} Radio (1.4\,GHz) contours (green) starting at $3\sigma$ and increasing in
steps of $\sigma = 9.5\,\mu$Jy. The thick black circle is the $6\arcsecsgreve$ search
radius adopted.} 
\label{fig:number-counts-and-follow-up}
\end{figure}

The 1200\,$\mu$m number counts derived from the survey are shown in Fig.~\ref{fig:number-counts-and-follow-up}a
(Greve et al.~in prep.).
At flux levels $\lsgreve 3.5$\,mJy the power-law slope of the number counts is about $\alpha \sim -1.6$, while 
at the brighter end there is evidence 
for a turn-over in the number counts, as is illustrated by 
the fact that the data are well matched by an integrated Schechter function with a knee at 3.5\,mJy. 
At a redshift of 2.5, this corresponds to a far-IR luminosity
of $10^{13}\,\Lsolargreve$ assuming a modified black body law with $\beta=1.5$ and $T_{d} = 40$\,K.
For comparison we have also plotted the 850\,$\mu$m counts from the HDF-N SCUBA Supermap \cite{Borys-et-al-2003},
scaled by a factor of $S_{850\mu\rm{m}}/S_{1200\mu\rm{m}} = 2.5$ which is expected 
for a starburst galaxy at $z=2.5$ \cite{Eales-et-al-2003}. Even though this scaling-factor is
highly uncertain, the agreement between the
1200\,$\mu$m and scaled 850\,$\mu$m counts in terms of the shape of the number counts is remarkably good.

Deep radio observations currently provide the most efficient way of determining
the exact positions of (sub)-mm sources, and thus positively identifying them in the optical/NIR
\cite{Ivison-et-al-1998,Smail-et-al-2000}. Using deep Very Large Array radio maps \cite{Ivison-et-al-2002}
we have searched for statistically robust radio counterparts within $6\arcsecsgreve$ of 
each of the MAMBO sources in our sample. 
We find that about two-thirds of the MAMBO sources 
have counterparts in the radio, which is comparable to the radio-identification fraction found
for SCUBA sources \cite{Ivison-et-al-2002}. 
The MAMBO source shown in Fig.~\ref{fig:number-counts-and-follow-up}b 
is associated with a very strong radio counterpart ($S_{1.4\,GHz} = 189\,\mu$Jy) which
lies on top of a compact optical/NIR galaxy. A Keck LRIS-B spectrum of this
source reveals that it is a type II QSO at $z=2.6$ (Ivison et al.~in prep.).
This source lies well within the SCUBA map yet is not included in the $\geq 3.0\sigma$ SCUBA catalogue 
\cite{Scott-et-al-2002}. While it is conceivable that a certain fraction of the 1200\,$\mu$m sources
might be at extremely high redshifts ($z > 8$) and thus can 'drop-out' at 850\,$\mu$m if the
dust is cold \cite{Eales-et-al-2003}, it is
clearly not the case here since it is detected in the I-band which is shortward of
912\,{\AA} for $z > 8$. Comparing the MAMBO and SCUBA maps we find that, although a few MAMBO 
sources are not detected by SCUBA and vice versa, there is a fair overall correlation between
the 1200\,$\mu$m and 850\,$\mu$m counts and galaxy positions, suggesting that both surveys are tracing the same high-redshift
dusty population (Greve et al.~in prep.). If this is the case, the faster mapping speed (about a factor of $\times 6$) and
smaller beam size of IRAM 30m/MAMBO over that of JCMT/SCUBA make the former the facility of choice
for wide-field extragalactic surveys.

%INDEX%%%%%%%%%%%%%%%%%%%%%%%%%%%%%%%%%%%%%%%%%%%%%%%%%%%%%%%%%%%%%%%
% Please check with the editor of your book whether he plans to
% include a "mutual" subject index - if so, please code your entries
% in the standard syntax. For your own purposes you may print your
% "personal" index by using the following commands:
%
%\clearpage
%\addcontentsline{toc}{section}{Index}
%\flushbottom
%\printindex
%%%%%%%%%%%%%%%%%%%%%%%%%%%%%%%%%%%%%%%%%%%%%%%%%%%%%%%%%%%%%%%%%%%%%

\end{document}